\documentclass[aps,pra,amsmath,amsfonts,amssymb,twocolumn,nofootinbib]{revtex4}

\usepackage{amssymb}
\usepackage{graphicx,psfrag,color}
\usepackage{dcolumn}
\usepackage{bm}
\usepackage{mathbbol}
\usepackage[T1]{fontenc}

\usepackage{amsthm}

\begin{document}
\flushbottom

\title{Teleportation-based quantum state tomography}

\author{Gustavo Rigolin}
\email{rigolin@ufscar.br}
\affiliation{Departamento de F\'isica, Universidade Federal de
S\~ao Carlos, 13565-905, S\~ao Carlos, SP, Brazil}

\date{\today}

\begin{abstract}
We explicitly show that the quantum teleportation protocol can be employed to completely reconstruct arbitrary two- and three-qubit density matrices. We also extend the present analysis to $n$-qubit density matrices. The only quantum resources needed to implement the teleportation-based quantum state tomography protocol
are the ability to make Bell measurements and the ability to prepare
a few different single qubit states to be teleported from Alice to Bob.  
\end{abstract}

\maketitle

\section{Introduction} 

The main goal of the original quantum teleportation protocol is to transfer the quantum state describing a single qubit in one location (Alice) onto another qubit in a 
different location (Bob) \cite{ben93,vai94,bra98,bow97,bos98,fur98}. The qubit to be teleported (actually its ``wave function'') is not physically sent from Alice to Bob and one does not even need to know its wave function for the protocol to work. The key ingredient needed to its implementation is a maximally entangled state shared between Alice and Bob (a two-qubit state), usually called the entangled resource or the quantum communication channel. They both must know which entangled state they share before starting the protocol to properly implement it. In addition to the 
entangled resource, Alice must be able to perform Bell measurements and Bob must be 
able to realize single qubit unitary operations at his share of the entangled resource to finish the protocol. These unitary ``corrections'' depend on which maximally entangled state Alice and Bob share as well as on the Bell measurement (BM) 
result obtained by Alice. As such, Alice
and Bob should have access to a reliable classical communication channel, through which Alice informs Bob of her BM result (two bits of classical information).  If all went well, at the end of one run of the protocol Bob's qubit (his share of the two-qubit entangled state) will be exactly described by the state teleported by Alice (input). Alice's input state, on the other hand, will be described by a maximally mixed state. No vestige of the original quantum state describing the input state remains at Alice's.

Let us assume now that we do not know the quantum resource shared by Alice and Bob but do know the input state to be teleported. Can we apply the teleportation protocol to reconstruct the two-qubit state shared between Alice and Bob? Can this be done by 
measuring only the single qubit with Bob after the execution of the teleportation protocol? 
What is the minimum number of different known 
input states that Alice needs to teleport to Bob to fully reconstruct the shared state? Can this be extended to reconstruct a multipartite quantum state?

In this work we address all the previous questions, presenting what we call the 
teleportation-based quantum state tomography (QST) protocol. We explicitly show that arbitrary
two- and three-qubit density matrices can be fully reconstructed using the teleportation protocol. 
We also show how to extend the present protocol to reconstruct
an arbitrary $n$-qubit density matrix, investigating the minimal number of known quantum states needed to be teleported from Alice to Bob for the protocol to work.

The present protocol is a conceptually different tool aiming at the fully reconstruction
of arbitrary density matrices, complementing the seminal strategies
to deal with such a problem that do not make use of the teleportation
protocol \cite{vog89,leo95,whi99,jam01,ari01,bri04,moh06,moh07,moh08,cra10,gro10,tot10,chr12,bau13,bau13b,lan17,tor18}. These tools are broadly known as 
QST protocols. See, for instance, Refs. \cite{ton19,cze22} for recent reviews addressing the most important QST protocols. We should also 
mention that one of the main goals of this article is to present the key ideas 
leading to the teleportation-based QST protocol. We are not addressing whether 
the present approach is more efficient than the standard ones. We want to show that
the quantum teleportation protocol, whose operation is fundamentally based on Bell measurements and classical communication, can be tailored to reconstruct an arbitrary quantum state. The quantum state to be reconstructed being the state shared among all the participants in the teleportation protocol.  In other words,  by repurposing the quantum teleportation protocol, we can change a tool originally developed for transferring  unknown  quantum states around into one that reveals it. This exchange of the roles of what is known and
unknown in the quantum teleportation protocol is the basic insight  that led to the teleportation-based QST protocol.

\section{Two-qubit states}

\subsection{The teleportation-based QST protocol}

The quantum state shared between Alice and Bob, and which we want
to fully reconstruct, is not in general a pure state. Therefore, we
need to recast the teleportation protocol in the language of density
matrices \cite{rig15,pav23}, the appropriate objects to describe an arbitrary quantum state, be it a pure or a mixed state.

\begin{figure}[!ht]
\includegraphics[height=2.75cm,width=5.25cm]{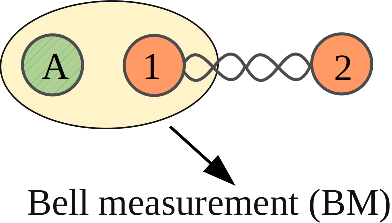}
\caption{\label{fig1}(color online) 
The teleportation-based QST starts
by Alice preparing an appropriate input state $\rho_A$. Subsequently,
she realizes a Bell measurement (BM) on her two qubits ($A$ and $1$), informing Bob of the obtained result as well as the state describing initially qubit $A$. With this information, Bob knows that his qubit is $\varrho_2$ [Eq.~(\ref{stepE})]. 
Repeating this process
for a few different inputs,
Bob can fully reconstruct
the initial state describing qubits $1$ and $2$, i.e. $\rho_{12}$,  with the knowledge of all states $\varrho_2$ associated with each different input state $\rho_A$.}
\end{figure}

The quantum state shared between Alice and Bob are labeled $1$ and $2$, respectively, while Alice's input state, i.e., the state
to be teleported, is labeled $A$ (see Fig.~\ref{fig1}). 
The density matrices describing those qubits are $\rho_{12}$ and
$\rho_{A}=|\psi\rangle\langle \psi|$. Alice's input will always be
a pure state, namely,
$|\psi\rangle=\alpha|0\rangle + \beta|1\rangle$, where
$|\alpha|^2+|\beta|^2=1$.

Initially, the state describing all qubits is 
\begin{equation}
\rho = \rho_{A} \otimes \rho_{12} = 
\left(
\begin{array}{cc}
|\alpha|^2 & \alpha\beta^*  \\
\alpha^*\beta & |\beta|^2
\end{array}
\right) \otimes \rho_{12},
\label{stepA}
\end{equation}
where $*$ denotes complex conjugation. After one complete run
of the protocol, Bob's qubit becomes \cite{rig15,pav23}
\begin{equation}
\varrho_{2}=(U_jTr_{A,1}[P_j \rho P_j]U_j^\dagger)/Q_j.
\label{stepD}
\end{equation}
In Eq.~(\ref{stepD}), $Tr_{A,1}$ is the partial trace on Alice's qubits, $j$ represents the Bell measurement (BM) result
obtained by Alice ($j=\Psi^-,\Psi^+,\Phi^-,\Phi^+$), and $P_j$ 
gives the projectors describing the four possible BM outcomes,
\begin{eqnarray}
P_{\Psi^{\pm}} = |\Psi^{\pm}\rangle \langle \Psi^{\pm}|,
&
P_{\Phi^{\pm}} = |\Phi^{\pm}\rangle \langle \Phi^{\pm}|, \label{projectorB}  
\end{eqnarray}
with the Bell states being
\begin{eqnarray}
|\Psi^{\pm}\rangle=(|01\rangle \pm |10\rangle)/\sqrt{2},
&|\Phi^{\pm}\rangle=(|00\rangle \pm |11\rangle)/\sqrt{2}. \label{BellB}
\end{eqnarray}

Alice's probability to measure a given Bell state $j$ depends on
the input state as well on the shared state and it is given by \cite{rig15,pav23}
\begin{equation}
Q_j=Q_j(\alpha,\beta) = Tr[{P_j \rho}],
\label{prob}
\end{equation}
where we have explicitly highlighted its dependence on the input
state for further convenience in the subsequent calculations.

The corrections that Bob must implement on his qubit after being informed of Alice's BM outcome is $U_j$. These unitary corrections 
are given by the identity operator and the standard Pauli matrices:
$\mathbb{1},\sigma^z,\sigma^x,\sigma^z\sigma^x$ \cite{nie00}. Since 
in the teleportation-based QST 
we always know the state to be teleported, those corrections are not needed and we can set all of them to $\mathbb{1}$. Therefore, at the end of each run of the teleportation protocol, Bob's state will be
given by 
\begin{equation}
\varrho_{2} =  \frac{Tr_{A,1}[P_j \rho P_j]}{Q_j}
=\frac{1}{Q_j}\left(
\begin{array}{cc}
\tilde{b}_{11} & \tilde{b}_{12}  \\
\tilde{b}_{12}^* & \tilde{b}_{22}
\end{array}
\right)
=\left(
\begin{array}{cc}
b_{11} & b_{12}  \\
b_{12}^* & b_{22}
\end{array}
\right). 
\label{stepE}
\end{equation}
Equation (\ref{stepE}) is the key result we need to build the teleportation-based QST protocol. Bob is supposed to be able to experimentally determine the matrix elements $b_{ij}$ of his single qubit 
state. They are functions of the shared state $\rho_{12}$ matrix elements as well as of the particular input state teleported by Alice. As we will see, by properly working with only four different inputs, we
will obtain a complete system of linear equations on the matrix elements of $\rho_{12}$ that allows us
to express them in terms of the matrix elements $b_{ij}$ of Bob's state. 
It is also convenient to write 
$\tilde{b}_{ij}=\tilde{b}_{ij}(\alpha,\beta)$, where
we highlight its dependence on the input state.

Since any physical density matrix must be positive definite and Hermitian, the most general two-qubit state, entangled or not, that we want to reconstruct can be written as follows:
\begin{equation}
\rho_{12}  = \left(
\begin{array}{cccc}
m_{11} & m_{12} & m_{13} & m_{14}\\
m_{12}^* & m_{22}  & m_{23} & m_{24} \\
m_{13}^* & m_{23}^* & m_{33} &  m_{34}\\
m_{14}^* & m_{24}^* & m_{34}^* & m_{44}
\end{array}
\right), 
\label{rho12}
\end{equation}
Note that $m_{jj}$ are all positive reals and $\sum_{j}m_{jj}=1$, with the latter being the normalization condition. We have a total of $16$ real parameters to determine ($15$ if we use the normalization condition): the real and imaginary parts of
$m_{ij}$, if $i \neq j$, and the four real quantities given by
$m_{jj}$.

For definiteness, and without loss of generality, we 
assume that Alice's BM result is always the Bell state $|\Psi^-\rangle$. The same analysis applies if she obtains any
other of the three remaining Bell states. It will only change the way Bob's coefficients $b_{ij}$ are related to the coefficients $m_{ij}$. See Appendix
\ref{apA} for more details. In all four cases, though, we will always have a compatible system of linear equations allowing us to solve all $m_{ij}$ as functions of $b_{ij}$. 

Using Eqs.~(\ref{stepA}), (\ref{projectorB}), and (\ref{rho12}),
it is not difficult to see that Alice's probability to measure 
the Bell state $|\Psi^-\rangle$ is
\begin{eqnarray}
Q_{\Psi^{\mbox{\small{-}}}}(\alpha,\beta) &=&[(m_{33}+m_{44})|\alpha|^2 + (m_{11}+m_{22})|\beta|^2]/2 \nonumber \\
&&-\mbox{Re}[(m_{13}+m_{24})\alpha^*\beta]. \label{AliceProb}
\end{eqnarray}
And if we also employ Eq.~(\ref{stepE}), we obtain
\begin{eqnarray}
\hspace{-0.5cm}\tilde{b}_{11}(\alpha,\beta) 
\hspace{-0.08cm}&\hspace{-0.08cm} =\hspace{-0.08cm}&\hspace{-0.08cm} 
(m_{33}|\alpha|^2 +m_{11}|\beta|^2)/2 
-\mbox{Re}[m_{13}\alpha^*\beta], \label{eqb11} \\
\hspace{-0.5cm}\tilde{b}_{22}(\alpha,\beta) 
\hspace{-0.08cm}&\hspace{-0.08cm} =\hspace{-0.08cm}&\hspace{-0.08cm} 
(m_{44}|\alpha|^2 +m_{22}|\beta|^2)/2 
-\mbox{Re}[m_{24}\alpha^*\beta], \label{eqb22} \\
\hspace{-0.5cm}2\tilde{b}_{12}(\alpha,\beta)
\hspace{-0.08cm}&\hspace{-0.08cm} =\hspace{-0.08cm}&\hspace{-0.08cm}
m_{34}|\alpha|^2 \!+\!m_{12}|\beta|^2 
\!-\!m_{14}\alpha^*\beta \!-\! m_{23}^*\alpha\beta^*\!.\label{eqb12}
\end{eqnarray}

Looking at Eqs.~(\ref{eqb11})-(\ref{eqb12}), we realize that if 
Alice teleports the state $|\psi\rangle=|0\rangle$ 
($\alpha,\beta=1,0$), we can determine 
the coefficients $m_{11},m_{22}$, and $m_{12}$. Analogously, if she
teleports the state $|\psi\rangle=|1\rangle$ ($\alpha,\beta=0,1$) we
obtain $m_{33},m_{44}$, and $m_{34}$. Finally, teleporting the states 
$|\psi\rangle=(|0\rangle+|1\rangle)/\sqrt{2}$  
($\alpha,\beta=1/\sqrt{2},1/\sqrt{2}$) and 
$|\psi\rangle=(|0\rangle+i|1\rangle)/\sqrt{2}$  
($\alpha,\beta=1/\sqrt{2},i/\sqrt{2}$), we arrive at a system
of linear equations on the real and imaginary parts of the remaining
coefficients ($m_{13},m_{14},m_{23},m_{24}$) that allows us to express them in terms of the previous coefficients and of the matrix elements of Bob's states associated with the latter two teleported states by Alice. Note that Bob's states are obtained experimentally (measured) by him.\footnote{Similarly to any QST protocol, several runs of the present scheme should be implemented for each state teleported by Alice. In this way Bob will have enough data to accurately reconstruct his qubit.} The matrix elements of his
states are the piece of data needed to fully reconstruct the two-qubit state shared between Alice and Bob. The solution to all those equations can be written as follows:
\begin{eqnarray}
m_{11} &=& 2 \tilde{b}_{11}(0, 1),\;\; m_{22} = 2 \tilde{b}_{22}(0, 1), \label{1of16}\\
m_{33} &=& 2 \tilde{b}_{11}(1, 0),\;\; m_{44} = 2 \tilde{b}_{22}(1, 0), \\
m_{12} &=& 2 \tilde{b}_{12}(0, 1),\;\; m_{34} = 2 \tilde{b}_{12}(1, 0),\\
m_{13} &=& (1 - i) [\tilde{b}_{11}(0, 1) + \tilde{b}_{11}(1, 0)] 
\nonumber \\
&+& 2 i \tilde{b}_{11}(1/\sqrt{2}, i/\sqrt{2})
-  2 \tilde{b}_{11}(1/\sqrt{2}, 1/\sqrt{2}), \\
m_{24} &=&(1 - i) [\tilde{b}_{22}(0, 1) + \tilde{b}_{22}(1, 0)] 
\nonumber \\
&+& 2 i \tilde{b}_{22}(1/\sqrt{2}, i/\sqrt{2})
-  2 \tilde{b}_{22}(1/\sqrt{2}, 1/\sqrt{2}), 
\end{eqnarray}
\begin{eqnarray}
m_{14} &=&(1 - i) [\tilde{b}_{12}(0, 1) + \tilde{b}_{12}(1, 0)] 
\nonumber \\
&+& 2 i \tilde{b}_{12}(1/\sqrt{2}, i/\sqrt{2})
-  2 \tilde{b}_{12}(1/\sqrt{2}, 1/\sqrt{2}), \\
m_{23} &=& (1 - i) [\tilde{b}_{12}^*(0, 1) + \tilde{b}_{12}^*(1, 0)] 
\nonumber \\
&+& 2 i \tilde{b}_{12}^*(1/\sqrt{2}, i/\sqrt{2})
-  2 \tilde{b}_{12}^*(1/\sqrt{2}, 1/\sqrt{2}).\label{16of16}
\end{eqnarray}

Some important remarks are in order now. First, the four states
teleported by Alice to Bob leading to the full reconstruction of 
the shared state between them are not unique. For instance, instead of the states
$|\psi\rangle=(|0\rangle+|1\rangle)/\sqrt{2}$  and 
$|\psi\rangle=(|0\rangle+i|1\rangle)/\sqrt{2}$, she can use the states $|\psi\rangle=(|0\rangle-|1\rangle)/\sqrt{2}$  and 
$|\psi\rangle=(|0\rangle-i|1\rangle)/\sqrt{2}$. The key point here is
that we must use at least four states. Each state leads to a different density matrix describing Bob's qubit at the end of the teleportation protocol. And since each density matrix possesses four 
independent real parameters (the two real diagonal entries and the complex non-diagonal one), four different teleported states will lead to $4\times 4 = 16$ real independent parameters. These $16$ independent parameters with Bob match exactly the number of real 
parameters needed to completely describe a two-qubit state. That is 
why we obtained $16$ linear equations in the real parameters of the
two-qubit state that we wanted to reconstruct and whose solutions 
are compactly expressed in Eqs.~(\ref{1of16})-(\ref{16of16}). Also,
we will show next that the extension of the previous analysis to
an $n$-qubit state is what leads to the generalization of the present
protocol to multipartite qubit systems.

Second, since we are assuming that Bob, and eventually Alice, can fully determine single qubit states, they can use this ability to 
determine some of the parameters of the two-qubit state without 
using the teleportation-based QST protocol. 
This is accomplished by measuring $\rho_1$ and $\rho_2$, 
the single qubit states obtained by tracing out 
one or the other qubit from $\rho_{12}$.

Third, the previous solution, Eqs.~(\ref{1of16})-(\ref{16of16}), is
also valid for low rank matrices. For example, if the rank of 
$\rho_{12}$ is only three, such that the fourth row and column are all zero, Eqs.~(\ref{1of16})-(\ref{16of16}) are still valid. 

\subsection{The standard QST protocol}

Let us now briefly compare the present approach with the usual way of reconstructing a two-qubit state \cite{jam01,ton19}. Our goal here is to illustrate the main conceptual differences between both strategies. In Ref. \cite{jam01},  a two-qubit density matrix is reconstructed by measuring the expectation values associated with a set of projectors that span the two-qubit Hilbert space. Being more explicit, we need to be able to make $16$
projective measurements. These $16$ projective measurements are given by $\Pi_j\otimes \Pi_k$, where $j,k=0,1,2,3$, and $\Pi_j$ are the following
single qubit projectors: $\Pi_0=|0 \rangle\langle 0|,\Pi_1=|1 \rangle\langle 1|,\Pi_2=|D \rangle\langle D|,
\Pi_3=|R \rangle \langle R|$. Here $|D \rangle=(|0\rangle+|1\rangle )/\sqrt{2}$ and
$|R \rangle=(|0\rangle-i|1\rangle)/\sqrt{2}$. This set of projectors is not unique and the ones just reported are usually employed in optical
experiments \cite{jam01}. Conceptually, the standard QST protocol probes the unknown two-qubit state from different local bases. The collected probabilities provide enough information to determine all independent parameters of the two-qubit density matrix. Once these probabilities are known,
the density matrix is reconstructed through a linear inversion procedure.

In the teleportation-based QST protocol, the goal is the same. We need to obtain a complete set of probabilities from which the density matrix can be fully reconstructed. The difference is that the information is extracted through local Bell state measurements rather than through local projective measurements in different bases. Typically, BMs with ancillary states are combined with all the steps of the teleportation protocol. These ingredients convert information about the unknown state into BM outcomes. The resulting probabilities play the same role as the coincidence probabilities in the standard protocol, giving a tomographically complete data set from which the density matrix can be reconstructed.

The fundamental distinction between both approaches lies in the measurement primitive employed to access the information contained in the unknown state. In the standard QST protocol,  tomography is achieved through a collection of local projective measurements performed in several complementary bases directly onto the two qubits. In the teleportation-based QST protocol, the same information is extracted through Bell state projections using the qubits of the unknown state to be reconstructed and the four ancillary states previously described.  The conceptual novelty of the teleportation-based QST protocol is therefore not in the mathematical reconstruction itself, but in demonstrating that the entire tomographic information can be acquired using BMs as the sole measurement resource, with the information associated with the unknown state teleported to a single qubit, the one with Bob.
See also Sec. \ref{oqs} for further details showing why only BMs are enough to fully reconstruct a two-qubit state.

In other words, the reconstruction stage follows the same philosophy as the standard QST protocol: a tomographically complete set of probabilities is first obtained and then linearly inverted to recover the density matrix. The difference lies in how these probabilities are acquired and ultimately where the parameters characterizing the unknown state are located. Whereas the standard QST protocol relies on local projective measurements in multiple bases,
with the information associated with the unknown state spread into the unknown two-qubit state, the teleportation-based QST protocol extracts the same tomographic information exclusively from Bell state measurement outcomes, with the information related to the unknown two-qubit state being transferred to Bob's qubit [see Eq.~(\ref{stepE}) and Eqs.~(\ref{eqb11})-(\ref{eqb12})].

\section{One-qubit states}
\label{oqs}

We can adapt the present protocol to reconstruct
a single qubit state. In other words, the  
ability to implement Bell measurements (BMs) and to prepare those four states, namely, $|0\rangle,|1\rangle,(|0\rangle+|1\rangle)/\sqrt{2}$,
 and $(|0\rangle+i|1\rangle)/\sqrt{2}$, suffices to reconstruct a 
single qubit state.
Moreover, by using the two-qubit QST protocol discussed above or the general $n$-qubit protocol presented later, we see that an arbitrary $n$-qubit state can be fully reconstructed using BMs alone, provided that the four states introduced above can be prepared and that the outcomes of all BMs performed by Alice are communicated classically to Bob. Indeed, once the teleportation-based QST protocol is executed, the only remaining task on Bob's side is the determination of single-qubit states. Crucially, this step does not require any measurement beyond BMs, since each single-qubit state can itself be reconstructed using the Bell-state-measurement-only single qubit reconstruction protocol described next. Therefore, every stage of the reconstruction procedure can be carried out exclusively through BMs, with no need for any other type of measurement.

Let an arbitrary single qubit be given by 
\begin{equation}
\varrho_{1} =  
\left(
\begin{array}{cc}
a_{11} & a_{12}  \\
a_{12}^* & a_{22}
\end{array}
\right). 
\label{stepF}
\end{equation}
The analogous to Eq.~(\ref{stepA}) is now $\rho=\rho_A\otimes \varrho_1$. If we make a Bell measurement on the two-qubit state 
$\rho$, the probability to obtain the Bell state $|\Psi^-\rangle$
is
\begin{equation}
Q_{\Psi^{\mbox{\small{-}}}}(\alpha,\beta) = 
[a_{22}|\alpha|^2 + a_{11}|\beta|^2]/2 
-\mbox{Re}[a_{12}\alpha^*\beta].
\label{qsingle}
\end{equation}

Looking at Eq.~(\ref{qsingle}), we realize that if $\rho_A$ 
is the density matrix representing the state $|0\rangle$, we obtain $a_{22}$ as a function of the probability to measure the Bell state
$|\Psi^-\rangle$. Similarly, using the state $|1\rangle$ gives
$a_{11}$ and using both $(|0\rangle + |1\rangle)/\sqrt{2}$ and
$(|0\rangle + i|1\rangle)/\sqrt{2}$ gives $a_{12}$. Writing the 
solution explicitly we have
\begin{eqnarray}
a_{11} &=& 2 Q_{\Psi^{\mbox{\small{-}}}}(0,1), \, 
a_{22} = 2 Q_{\Psi^{\mbox{\small{-}}}}(1,0), \\ 
a_{12} &=& (1-i)[Q_{\Psi^{\mbox{\small{-}}}}(1,0)+Q_{\Psi^{\mbox{\small{-}}}}(0,1)] \nonumber \\
&+&\hspace{-.1cm} 
2iQ_{\Psi^{\mbox{\small{-}}}}(1/\sqrt{2},i/\sqrt{2})
-2Q_{\Psi^{\mbox{\small{-}}}}(1/\sqrt{2},1/\sqrt{2}).
\label{single}
\end{eqnarray}
Before we move on to the three-qubit state protocol, we should 
mention that the choice for $|\Psi^-\rangle$ as the BM
result above is just for convenience and definiteness.
Similar results follow for the three other possible 
outcomes of a BM. There is nothing particularly special about the Bell state $|\Psi^-\rangle$ when it comes to the
quantum state reconstruction protocols being studied here.

\section{Three-qubit states}

\begin{figure}[!ht]
\includegraphics[height=2.75cm,width=5.25cm]{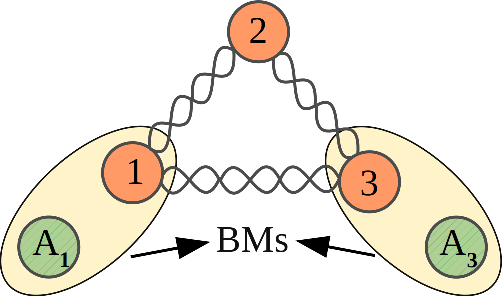}
\caption{\label{fig2}(color online) 
The three-qubit teleportation-based QST. Alice prepares two input states ($\rho_{A_1}$ and
$\rho_{A_3}$) and then realizes independent BMs on each pair 
of qubits as shown in the figure, informing Bob of the obtained results and the states  describing initially qubits $A_1$ and $A_3$.
With this information, Bob knows that his qubit is given by 
$\varrho_2$ [Eq.~(\ref{stepE3})]. 
Repeating this process
for a few different combinations of pairs of input states 
$\rho_{A_1}$ and $\rho_{A_3}$, Bob can reconstruct
the initial state describing $\rho_{123}$ if he knows all the 
states $\varrho_2$ associated with each different pair of input states.
}
\end{figure}

The three-qubit telepor\-tation-based
QST is very similar to the two-qubit case. Most of the equations are formally the same. The main changes are given by Eq.~(\ref{stepA}),  which now becomes
\begin{eqnarray}
\rho &=& \rho_{A_{1}} \otimes \rho_{123} \otimes \rho_{A_{3}} \nonumber \\
&=& 
\left(
\begin{array}{cc}
|\alpha|^2 & \alpha\beta^*  \\
\alpha^*\beta & |\beta|^2
\end{array}
\right) \otimes \rho_{123} \otimes
\left(
\begin{array}{cc}
|\gamma|^2 & \gamma\delta^*  \\
\gamma^*\delta & |\delta|^2
\end{array}
\right).
\label{stepA3}
\end{eqnarray}
In other words, Alice now teleports a pair of input states,
namely, $|\psi\rangle_{A_1}=\alpha|0\rangle + \beta|1\rangle$
and $|\psi\rangle_{A_3}=\gamma|0\rangle + \delta|1\rangle$, 
where $|\alpha|^2+|\beta|^2=1$ and $|\gamma|^2+|\delta|^2=1$ 
(see Fig. \ref{fig2}).

After Alice implements a BM on qubits $1$ and $A_1$, another one
on qubits $3$ and $A_3$, and tells Bob her results, Bob
qubit is given by an expression similar to Eq.~(\ref{stepE}),
where we now need two indexes to properly label the $16$ possible
outcomes of the two BMs implemented by Alice:
\begin{equation}
\varrho_{2} =  \frac{Tr_{A_1,1,A_3,3}[P_{i,j} \rho P_{i,j}]}{Q_{i,j}}
=\left(
\begin{array}{cc}
b_{11} & b_{12}  \\
b_{12}^* & b_{22}
\end{array}
\right). 
\label{stepE3}
\end{equation}
In Eq.~(\ref{stepE3}), $Q_{i,j}$ is Alice's probability to
measure Bell states $i$ and $j$ and $P_{i,j}$ is the projector
describing the two independent BMs. Being more explicit,
\begin{eqnarray}
Q_{i,j} &=& Tr[{P_{i,j} \rho}], \label{prob3} \\
P_{i,j} &=& P_{i} \otimes \mathbb{1} \otimes P_{j}, 
\end{eqnarray}
where $P_{i}$ and $P_{j}$  
are given by Eq.~(\ref{projectorB}).

Similarly to the two-qubit state protocol, we assume that Bob
can experimentally determine the matrix elements $b_{ij}$ of his single qubit state, which are now functions of the matrix elements of  $\rho_{123}$ and of the pair of input states teleported by Alice. As we show in Appendix \ref{apB}, 
using all $16$ pairs of inputs that can be formed out of  
the states $|0\rangle, |1\rangle,
(|0\rangle+|1\rangle)/\sqrt{2},$ and $(|0\rangle+i|1\rangle)/\sqrt{2}$, we can express the $64$ real parameters that define
$\rho_{123}$ in terms of the matrix elements $b_{ij}(\alpha,\beta,\gamma,\delta)$ of Bob's states. 

\section{$n$-qubit states}

The remarkable feature of the three-qubit protocol is that it 
works perfectly by Alice teleporting only separable pairs of qubits. There is no need to teleport entangled states (a
Bell state, for instance). Also, there is no need to implement
more sophisticated joint measurements involving all four
qubits that are with Alice ($1, A_1, 3, A_3$). For the 
three-qubit protocol, the $16$ possible choices of pairs of qubits 
to be teleported from Alice to Bob lead to $16$ different density matrices at Bob's, one different density matrix at the end of each one of the $16$ different teleportation protocols that can be implemented using a particular pair as input. We are considering, for definiteness, only the cases where Alice obtains 
$|\Psi^-\rangle_{A_1,1}|\Psi^-\rangle_{A_3,3}$ as the result of her 
two independent BMs. Each density matrix has four independent real parameters, leading to a total of $4\times 16$
real parameters with Bob. These $64$ real parameters are connected to $64$ linear equations that can be solved to give 
the $64$ real parameters defining $\rho_{123}$.
\begin{figure}[!ht]
\includegraphics[width=7cm]{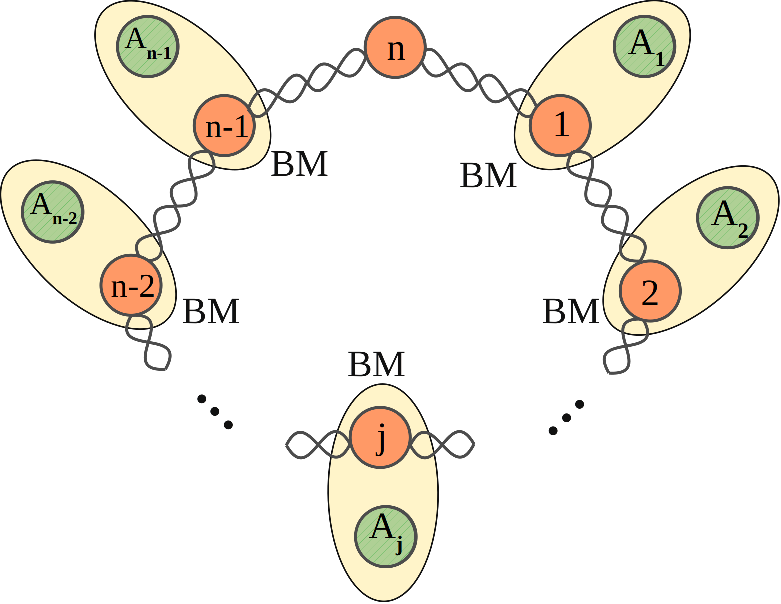}
\caption{\label{fig3}(color online) 
The $n$-qubit teleportation-based QST. Alice prepares $n-1$ input states ($\rho_{A_1}, \ldots, \rho_{A_{n-1}}$) and then realizes independent BMs on each pair 
of qubits as indicated in the figure, informing Bob of the results obtained and the states describing initially qubits 
$A_1, \ldots, A_{n-1}$.
With this information, Bob knows that his qubit (denoted by $n$
in the figure) is given by 
the generalized version of $\varrho_2$ [Eq.~(\ref{stepE3})]. 
Repeating this process
for several different arrangements of $n-1$ inputs, 
Bob can reconstruct
the initial state describing the $n$-qubit state 
if he knows all states $\varrho_2$ related to each different $n-1$
input state arrangement teleported by Alice.}
\end{figure}
If we now have an $n$-qubit state to be reconstructed, Alice needs to teleport $n-1$ single qubits, using 
as input all $n-1$ arrangements of states made out of the following
four qubits, namely, 
$|0\rangle, |1\rangle,
(|0\rangle+|1\rangle)/\sqrt{2},$ and 
$(|0\rangle+i|1\rangle)/\sqrt{2}$ (see Fig. \ref{fig3}).

Indeed, an $n$-qubit density matrix has $2^n\times 2^n$ elements.
The $n$ diagonal elements are real and the remaining ones are 
complex, with only half of the latter independent since the 
density matrix is Hermitian. There are, thus, a total of 
$4^n$ real parameters that we must know to fully reconstruct it. On the other hand, with four different 
single qubit states, we have a total of $4^{n-1}$ different
arrangements comprising $n-1$ qubits. Each one of 
these arrangements will lead to one different density matrix
with Bob at the end of the teleportation protocol
(after the $n-1$ BMs that Alice implements as described in Fig. \ref{fig3}).  As always, we are assuming, for definiteness, a 
particular sequence of BM results obtained by Alice (for instance, $|\Psi^-\rangle_{A_1,1}\otimes\cdots
\otimes|\Psi^-\rangle_{A_j,j}\otimes\cdots\otimes|\Psi^-\rangle_{A_{n-1},n-1}$), but the present discussion applies to 
any other sequence of BMs as well.
Since a single qubit density matrix provides four real parameters, Bob will end up with $4\times 4^{n-1} = 4^n$ real
parameters. This number matches exactly the number of independent
real parameters characterizing an $n$-qubit density matrices, 
allowing Bob to solve a system of $4^n$ linear equations that 
eventually gives all the matrix elements of the $n$-qubit 
state as functions of the matrix elements of all Bob's single
qubit density matrices.

\section{Conclusion}

We showed that the quantum teleportation protocol 
can be used to fully reconstruct an arbitrary $n$-qubit state. 
The teleportation-based quantum state tomography (QST) protocol 
here developed works by using the basic ``primitives'' of the teleportation protocol, namely,
Bell state measurements and classical communication. If the quantum states to be teleported from Alice to Bob are known, this knowledge
can be used to reconstruct the quantum state shared between them and
through which the teleportation occurs. We also showed that the minimum number of known states
that Alice must teleport is four and that they all lie on the equator and the poles of the Bloch sphere \cite{nie00}.

For the protocol to work, Alice should be able to prepare several 
copies of those four states and she should be able to realize independent Bell measurements (BMs). In other words, 
she needs to project 
those qubits and the ones coming from the state shared with Bob onto Bell states. Bob should be able to determine his single qubit at the end of the teleportation protocol. This can
be accomplished by any well-established  
QST protocol \cite{vog89,leo95,whi99,jam01,ari01,bri04,moh06,moh07,moh08,cra10,gro10,tot10,chr12,bau13,bau13b,lan17,tor18} or by simply using the Bell-state-measurement-only single qubit reconstruction protocol also 
discussed in this work. Each different
teleported state by Alice will lead to a different single qubit 
density matrix for Bob. These several density matrices with Bob 
contain all the data needed to reconstruct the shared state between
them. They provide us with a complete set of linear equations where the density matrix elements of the shared state are the 
unknown variables. By solving this set of equations, we obtain
the shared state density matrix elements in terms of the 
density matrix elements of Bob's single qubits. 

The teleportation-based QST protocol here developed works even if
the shared state is distributed among different parties located in different places (see Figs. \ref{fig2} and \ref{fig3}). If all parties cooperate, i.e., realize their BMs and 
communicate classically their results to Bob, he will be able to 
fully reconstruct the shared state by operating locally on his 
qubit as prescribed by the teleportation-based QST protocol.
Moreover, if we have some prior knowledge of the state shared among
all parties, the number of states needed to be teleported by Alice
and the resources needed to execute the protocol are reduced. An
interesting extension of this work would be the investigation of the
minimal resources needed to reconstruct matrix product states \cite{fan92,gar07} or nearly pure states, which require much less resources to be reconstructed using traditional QST techniques \cite{cra10,gro10,bau13,bau13b,lan17}. Also, the investigation of the possible connection between quantum data hiding \cite{div01,div02}
and local QST protocols may lead to a deeper understanding of the inherent limitations of tomography schemes based on local operations and classical communication.  

Finally, we would like to mention that the present protocol naturally
leads to the following two open problems. First, is it possible to
extend the present analysis for higher spins? For instance, can we
adapt the spin-1 teleportation protocol \cite{zei19} 
to fully reconstruct 
density matrices describing single and multipartite spin-1 systems?  
Second, can the continuous-variable (CV) teleportation protocol 
\cite{vai94,bra98,fur98} be modified to reconstruct CV-states, 
such as a two-mode or $n$-mode Gaussian state, where only local 
operations and classical communication are allowed in the 
reconstruction protocol \cite{har07,rig08}?

\textit{Note:} The extension of the present protocol to spin-1 systems is reported in Ref. \cite{rig26}.

\begin{acknowledgments}
GR thanks the Brazilian agency CNPq (National Council for
Scientific and Technological Development) for funding. 
\end{acknowledgments}

\appendix   

\section{The two-qubit case}
\label{apA}

If Alice obtains the Bell state $j$ after her Bell measurement,
the unnormalized state $\tilde{\varrho}_{2,j}$ with Bob is 
[see Eq.~(\ref{stepE})] 
\begin{equation}
\tilde{\varrho}_{2,j} = Tr_{A,1}[P_j \rho P_j]
=\left(
\begin{array}{cc}
\tilde{b}_{11,j} & \tilde{b}_{12,j}  \\
\tilde{b}_{12,j}^* & \tilde{b}_{22,j}
\end{array}
\right). 
\label{ustepE}
\end{equation}

Using Eqs.~(\ref{stepA}), (\ref{projectorB}), (\ref{rho12}),
and (\ref{ustepE}), we get that Alice's probability to measure 
the Bell state $j$ is
\begin{eqnarray}
Q_{\Psi^{\mbox{\small{-}}}}(\alpha,\beta) &=&[(m_{33}+m_{44})|\alpha|^2 +(m_{11}+m_{22})|\beta|^2]/2 \nonumber \\
&&-\mbox{Re}[(m_{13}+m_{24})\alpha^*\beta], \\
Q_{\Psi^{\mbox{\tiny{+}}}}(\alpha,\beta) &=& Q_{\Psi^{\mbox{\small{-}}}}(-\alpha,\beta), \\
Q_{\Phi^{\mbox{\small{-}}}}(\alpha,\beta) &=& Q_{\Psi^{\mbox{\small{-}}}}(\beta,\alpha) \\
Q_{\Phi^{\mbox{\tiny{+}}}}(\alpha,\beta) &=& Q_{\Psi^{\mbox{\small{-}}}}(-\beta,\alpha),
\label{AliceProb1}
\end{eqnarray}
and the corresponding Bob's unnormalized state is
\begin{eqnarray}
\tilde{\varrho}_{2,\Psi^{\mbox{\small{-}}}}(\alpha,\beta) &=& 
\left(
\begin{array}{cc}
\tilde{b}_{11,\Psi^{\mbox{\small{-}}}}(\alpha,\beta) & \tilde{b}_{12,\Psi^{\mbox{\small{-}}}}(\alpha,\beta)  \\
\tilde{b}_{12,\Psi^{\mbox{\small{-}}}}^*(\alpha,\beta) & \tilde{b}_{22,\Psi^{\mbox{\small{-}}}}(\alpha,\beta)
\end{array}
\right),\\
\tilde{\varrho}_{2,\Psi^{\mbox{\tiny{+}}}}(\alpha,\beta) &=&
\tilde{\varrho}_{2,\Psi^{\mbox{\small{-}}}}(-\alpha,\beta), \\
\tilde{\varrho}_{2,\Phi^{\mbox{\small{-}}}}(\alpha,\beta) &=&
\tilde{\varrho}_{2,\Psi^{\mbox{\small{-}}}}(\beta,\alpha), \\
\tilde{\varrho}_{2,\Phi^{\mbox{\tiny{+}}}}(\alpha,\beta) &=&
\tilde{\varrho}_{2,\Psi^{\mbox{\small{-}}}}(-\beta,\alpha), 
\end{eqnarray}
where
\begin{eqnarray*}
\hspace{-0.5cm}2\tilde{b}_{11,\Psi^{\mbox{\small{-}}}}(\alpha,\beta)
\hspace{-0.08cm}&\hspace{-0.08cm} =\hspace{-0.08cm}&\hspace{-0.08cm} 
m_{33}|\alpha|^2 +m_{11}|\beta|^2 
-2\mbox{Re}[m_{13}\alpha^*\beta], \label{eqb11a} \\
\hspace{-0.5cm}2\tilde{b}_{22,\Psi^{\mbox{\small{-}}}}(\alpha,\beta) 
\hspace{-0.08cm}&\hspace{-0.08cm} =\hspace{-0.08cm}&\hspace{-0.08cm} 
m_{44}|\alpha|^2 +m_{22}|\beta|^2 
-2\mbox{Re}[m_{24}\alpha^*\beta], \label{eqb22a} \\
\hspace{-0.5cm}2\tilde{b}_{12,\Psi^{\mbox{\small{-}}}}(\alpha,\beta) 
\hspace{-0.08cm}&\hspace{-0.08cm} =\hspace{-0.08cm}&\hspace{-0.08cm}
m_{34}|\alpha|^2 \!+\!m_{12}|\beta|^2 
\!-\!m_{14}\alpha^*\beta \!-\! m_{23}^*\alpha\beta^*\!.\label{eqb12a}
\end{eqnarray*}

\section{The three-qubit case}
\label{apB}

The three-qubit state that we want to reconstruct can be written as follows:
\begin{equation}
\rho_{123}  = \left(
\begin{array}{cccccccc}
m_{11} & m_{12} & m_{13} & m_{14} & m_{15} & m_{16} & m_{17} & m_{18}\\
m_{12}^* & m_{22}  & m_{23} & m_{24} & m_{25} & m_{26} & m_{27} & m_{28} \\
m_{13}^* & m_{23}^*  & m_{33} & m_{34} & m_{35} & m_{36} & m_{37} & m_{38}\\
m_{14}^* & m_{24}^* & m_{34}^* & m_{44} & m_{45} & m_{46} & m_{47} & m_{48} \\
m_{15}^* & m_{25}^* & m_{35}^* & m_{45}^* & m_{55} & m_{56} & m_{57} & m_{58} \\
m_{16}^* & m_{26}^* & m_{36}^* & m_{46}^* & m_{56}^* & m_{66} & m_{67} & m_{68} \\
m_{17}^* & m_{27}^* & m_{37}^* & m_{47}^* & m_{57}^* & m_{67}^* & m_{77} & m_{78} \\
m_{18}^* & m_{28}^* & m_{38}^* & m_{48}^* & m_{58}^* & m_{68}^* & m_{78}^* & m_{88}
\end{array}
\right), 
\label{rho123}
\end{equation}
where $m_{jj}$ are positive reals and $\sum_{j}m_{jj}=1$. We have  $64$ real parameters to determine ($63$ if we use the normalization condition): the real and imaginary parts of
$m_{ij}$, if $i \neq j$, and the eight real quantities given by
$m_{jj}$.

For definiteness, we 
assume that Alice's two BM results are always given by 
$|\Psi^-\rangle_{A_1,1}|\Psi^-\rangle_{A_3,3}$. The same analysis applies for any other of the remaining $15$ pairs of Bell states
that she can measure. Similar to what we showed for the two-qubit case (Appendix \ref{apA}),  we can obtain anyone of those other
cases by properly changing the signs and/or exchanging the values of 
$\alpha,\beta,\gamma,\delta$ 
in the expressions for the case we will be dealing with from now on.

\begin{widetext}
Using Eqs.~(\ref{stepA3}) and (\ref{prob3}), we can compute 
Alice's probability to measure 
the Bell states $|\Psi^-\rangle_{A_1,1}|\Psi^-\rangle_{A_3,3}$ if
she teleports the states $\alpha|0\rangle + \beta|1\rangle$ and
$\gamma|0\rangle + \delta|1\rangle$. This probability, given by
$Q_{\Psi^{\mbox{\small{-}}}\Psi^{\mbox{\small{-}}}}(\alpha,\beta,\gamma,\delta)$, is
\begin{eqnarray}
4Q_{\Psi^{\mbox{\small{-}}}\Psi^{\mbox{\small{-}}}}(\alpha,\beta,\gamma,\delta) &=& 
(m_{66}+m_{88})|\alpha \gamma|^2 + (m_{55}+m_{77})|\alpha \delta|^2
+ (m_{22}+m_{44})|\beta \gamma|^2 + (m_{11}+m_{33})|\beta \delta|^2
\nonumber \\
&&-2\mbox{Re}[(m_{56} + m_{78})\delta\gamma^*]|\alpha|^2
-2\mbox{Re}[(m_{12} + m_{34})\delta\gamma^*]|\beta|^2
-2\mbox{Re}[(m_{26} + m_{48})\beta\alpha^*]|\gamma|^2
\nonumber \\
&&-2\mbox{Re}[(m_{15} + m_{37})\beta\alpha^*]|\delta|^2
+2\mbox{Re}[(m_{16} + m_{38})\beta\delta\alpha^*\gamma^*]
+2\mbox{Re}[(m_{25} + m_{47})\alpha^*\delta^*\beta\gamma].
\label{AliceProb3}
\end{eqnarray}
If we also use Eq.~(\ref{stepE3}), 
the matrix elements of Bob's qubit at the end of 
the teleportation protocol are
\begin{equation}
b_{ij}(\alpha,\beta,\gamma,\delta)=
\tilde{b}_{ij}(\alpha,\beta,\gamma,\delta)/Q_{\Psi^{\mbox{\small{-}}}\Psi^{\mbox{\small{-}}}}(\alpha,\beta,\gamma,\delta), \label{btilde3}
\end{equation}
where $\tilde{b}_{ij}(\alpha,\beta,\gamma,\delta)$, 
the unnormalized coefficients, 
are
\begin{eqnarray}
4\tilde{b}_{11}(\alpha,\beta,\gamma,\delta) &=& 
m_{66}|\alpha\gamma|^2 +m_{55}|\alpha\delta|^2
+m_{22}|\beta \gamma|^2 +m_{11}|\beta \delta|^2
-2\mbox{Re}[m_{56}\delta\gamma^*]|\alpha|^2
-2\mbox{Re}[m_{12}\delta\gamma^*]|\beta|^2 \nonumber \\
&& -2\mbox{Re}[m_{26}\beta\alpha^*]|\gamma|^2
-2\mbox{Re}[m_{15}\beta\alpha^*]|\delta|^2
+2\mbox{Re}[m_{16}\beta\delta\alpha^*\gamma^*]
+2\mbox{Re}[m_{25}\alpha^*\delta^*\beta\gamma],
\label{eqb113}\\ 
4\tilde{b}_{22}(\alpha,\beta,\gamma,\delta) &=& 
m_{88}|\alpha\gamma|^2 + m_{77}|\alpha\delta|^2
+m_{44}|\beta \gamma|^2 + m_{33}|\beta \delta|^2
-2\mbox{Re}[m_{78}\delta\gamma^*]|\alpha|^2
-2\mbox{Re}[m_{34}\delta\gamma^*]|\beta|^2 \nonumber \\
&&-2\mbox{Re}[m_{48}\beta\alpha^*]|\gamma|^2
-2\mbox{Re}[m_{37}\beta\alpha^*]|\delta|^2
+2\mbox{Re}[m_{38}\beta\delta\alpha^*\gamma^*]
+2\mbox{Re}[m_{47}\alpha^*\delta^*\beta\gamma],
\label{eqb223}\\
4\tilde{b}_{12}(\alpha,\beta,\gamma,\delta) &=&  
m_{68}|\alpha\gamma|^2 + m_{57}|\alpha\delta|^2 
+ m_{24}|\beta\gamma|^2  + m_{13}|\beta\delta|^2 
-m_{58}\delta\gamma^* |\alpha|^2 -m_{67}\gamma\delta^* |\alpha|^2
- m_{14}\delta\gamma^* |\beta|^2
\nonumber \\
&& -m_{23}\gamma\delta^* |\beta|^2 
-m_{28}\beta \alpha^*|\gamma|^2 - m_{46}^*\alpha\beta^* |\gamma|^2
-m_{17}\beta\alpha^*|\delta|^2 - m_{35}^*\alpha\beta^*|\delta|^2
+ m_{18}\beta\delta\alpha^*\gamma^* 
\nonumber \\
&& + m_{27}\beta\gamma\alpha^*\delta^* 
+ m_{36}^*\alpha\gamma\beta^*\delta^* 
+ m_{45}^*\alpha\delta\beta^*\gamma^*. 
\label{eqb123}
\end{eqnarray}
%\end{widetext}

Looking at Eqs.~(\ref{eqb113})-(\ref{eqb123}), we notice that if 
Alice teleports the states $|0\rangle|0\rangle, |0\rangle|1\rangle, |1\rangle|0\rangle$,  
and $|1\rangle|1\rangle$, we can determine all diagonal elements of $\rho_{123}$ employing
the diagonal entries of Bob's states. The non-diagonal entries give the values of 
$m_{13},m_{24},m_{57}$, and $m_{68}$.
Now, if she teleports the pair of states 
$|0\rangle|+\rangle, |1\rangle|+\rangle, |+\rangle|0\rangle$,  
and $|+\rangle|1\rangle$, where $|+\rangle=(|0\rangle + |1\rangle)/\sqrt{2}$,
we get the real parts of $m_{12},m_{15},m_{26},m_{34},m_{37},m_{48},m_{56},m_{78}$
by working with the diagonal elements of Bob's states. The imaginary parts of those
eight coefficients are obtained dealing with the diagonal entries of Bob's states 
if Alice teleports the pair of states 
$|0\rangle|R\rangle, |1\rangle|R\rangle, |R\rangle|0\rangle$,  
and $|R\rangle|1\rangle$, where $|R\rangle=(|0\rangle + i|1\rangle)/\sqrt{2}$.
On the other hand, the previous two sets of non-diagonal terms give   
$m_{14},m_{17},m_{23},m_{28},m_{35}$, $m_{46},m_{58},m_{67}$. Finally, if Alice teleports
the following four pairs of states, namely, $|+\rangle|+\rangle, |+\rangle|R\rangle, 
|R\rangle|+\rangle$, and $|R\rangle|R\rangle$, we get the remaining coefficients, i.e.,
$m_{16},m_{18}$, $m_{25},m_{27},m_{38},m_{47},m_{36},m_{45}$.
Being more explicit, we can write all $m_{ij}$ as follows:
%
%\begin{widetext}
\begin{eqnarray}
m_{11} &=& 4 \tilde{b}_{11}(0, 1, 0, 1),\;\; m_{22} = 4 \tilde{b}_{22}(0, 1, 0, 1), \;\; m_{33} = 4 \tilde{b}_{11}(0, 1, 1, 0),\;\; m_{44} = 4 \tilde{b}_{22}(0, 1, 1, 0), \label{1of64}\\
m_{55} &=& 4 \tilde{b}_{11}(1, 0, 0, 1),\;\; m_{66} = 4 \tilde{b}_{22}(1, 0, 0, 1), \;\; m_{77} = 4 \tilde{b}_{11}(1, 0, 1, 0),\;\; m_{88} = 4 \tilde{b}_{22}(1, 0, 1, 0),\\
m_{13} &=& 4 \tilde{b}_{12}(0, 1, 0, 1),\;\; m_{24} = 4 \tilde{b}_{12}(0, 1, 1, 0),\;\; m_{57} = 4 \tilde{b}_{12}(1, 0, 0, 1), \;\; m_{68} = 4 \tilde{b}_{12}(1, 0, 1, 0),\\ 
m_{12} &=& 2(1 - i) [\tilde{b}_{11}(0, 1, 0, 1) + 
\tilde{b}_{11}(0, 1, 1, 0)] - 4 [ \tilde{b}_{11}(0, 1, 1/\sqrt{2}, 1/\sqrt{2}) - i \tilde{b}_{11}(0, 1, 1/\sqrt{2}, i/\sqrt{2})], \\
m_{34} &=& 2(1 - i) [\tilde{b}_{22}(0, 1, 0, 1) + 
\tilde{b}_{22}(0, 1, 1, 0)] - 4 [ \tilde{b}_{22}(0, 1, 1/\sqrt{2}, 1/\sqrt{2}) - i \tilde{b}_{22}(0, 1, 1/\sqrt{2}, i/\sqrt{2})], \\
m_{56} &=& 2(1 - i) [\tilde{b}_{11}(1, 0, 0, 1) + 
\tilde{b}_{11}(1, 0, 1, 0)] - 4 [ \tilde{b}_{11}(1, 0, 1/\sqrt{2}, 1/\sqrt{2}) - i \tilde{b}_{11}(1, 0, 1/\sqrt{2}, i/\sqrt{2})], \\
m_{78} &=& 2(1 - i) [\tilde{b}_{22}(1, 0, 0, 1) + 
\tilde{b}_{22}(1, 0, 1, 0)] - 4 [ \tilde{b}_{22}(1, 0, 1/\sqrt{2}, 1/\sqrt{2}) - i \tilde{b}_{22}(1, 0, 1/\sqrt{2}, i/\sqrt{2})], \\
m_{15} &=& 2(1 - i) [\tilde{b}_{11}(0, 1, 0, 1) + 
\tilde{b}_{11}(1, 0, 0, 1)] - 4 [ \tilde{b}_{11}(1/\sqrt{2}, 1/\sqrt{2},0, 1) - i \tilde{b}_{11}(1/\sqrt{2}, i/\sqrt{2},0, 1)], \\
m_{37} &=& 2(1 - i) [\tilde{b}_{22}(0, 1, 0, 1) + 
\tilde{b}_{22}(1, 0, 0, 1)] - 4 [ \tilde{b}_{22}(1/\sqrt{2}, 1/\sqrt{2}, 0, 1) - i \tilde{b}_{22}(1/\sqrt{2}, i/\sqrt{2}, 0, 1)], \\
m_{26} &=& 2(1 - i) [\tilde{b}_{11}(0, 1, 1, 0) + 
\tilde{b}_{11}(1, 0, 1, 0)] - 4 [ \tilde{b}_{11}(1/\sqrt{2}, 1/\sqrt{2}, 1, 0) - i \tilde{b}_{11}(1/\sqrt{2}, i/\sqrt{2}, 1, 0)], \\
m_{48} &=& 2(1 - i) [\tilde{b}_{22}(0, 1, 1, 0) + 
\tilde{b}_{22}(1, 0, 1, 0)] - 4 [ \tilde{b}_{22}(1/\sqrt{2}, 1/\sqrt{2}, 1, 0) - i \tilde{b}_{22}(1/\sqrt{2}, i/\sqrt{2}, 1, 0)], \\
m_{17} &=& 2(1 - i) [\tilde{b}_{12}(0, 1, 0, 1) + 
\tilde{b}_{12}(1, 0, 0, 1)] - 4 [ \tilde{b}_{12}(1/\sqrt{2}, 1/\sqrt{2}, 0, 1) - i \tilde{b}_{12}(1/\sqrt{2}, i/\sqrt{2}, 0, 1)], \\
m_{35} &=& 2(1 - i) [\tilde{b}_{12}^*(0, 1, 0, 1) + 
\tilde{b}_{12}^*(1, 0, 0, 1)] - 4 [ \tilde{b}_{12}^*(1/\sqrt{2}, 1/\sqrt{2}, 0, 1) - i \tilde{b}_{12}^*(1/\sqrt{2}, i/\sqrt{2}, 0, 1)], \\
m_{28} &=& 2(1 - i) [\tilde{b}_{12}(0, 1, 1, 0) + 
\tilde{b}_{12}(1, 0, 1, 0)] - 4 [ \tilde{b}_{12}(1/\sqrt{2}, 1/\sqrt{2}, 1, 0) - i \tilde{b}_{12}(1/\sqrt{2}, i/\sqrt{2}, 1, 0)], \\
m_{46} &=& 2(1 - i) [\tilde{b}_{12}^*(0, 1, 1, 0) + 
\tilde{b}_{12}^*(1, 0, 1, 0)] - 4 [ \tilde{b}_{12}^*(1/\sqrt{2}, 1/\sqrt{2}, 1, 0) - i \tilde{b}_{12}^*(1/\sqrt{2}, i/\sqrt{2}, 1, 0)], \\
m_{58} &=& 2(1 - i) [\tilde{b}_{12}(1, 0, 0, 1) + 
\tilde{b}_{12}(1, 0, 1, 0)] - 4 [ \tilde{b}_{12}(1, 0, 1/\sqrt{2}, 1/\sqrt{2}) - i \tilde{b}_{12}(1, 0, 1/\sqrt{2}, i/\sqrt{2})], \\
m_{67} &=& 2(1 + i) [\tilde{b}_{12}(1, 0, 0, 1) + 
\tilde{b}_{12}(1, 0, 1, 0)] - 4 [ \tilde{b}_{12}(1, 0, 1/\sqrt{2}, 1/\sqrt{2}) + i \tilde{b}_{12}(1, 0, 1/\sqrt{2}, i/\sqrt{2})], \\
m_{14} &=& 2(1 - i) [\tilde{b}_{12}(0, 1, 0, 1) + 
\tilde{b}_{12}(0, 1, 1, 0)] - 4 [ \tilde{b}_{12}(0, 1, 1/\sqrt{2}, 1/\sqrt{2}) - i \tilde{b}_{12}(0, 1, 1/\sqrt{2}, i/\sqrt{2})], \\
m_{23} &=& 2(1 + i) [\tilde{b}_{12}(0, 1, 0, 1) + 
\tilde{b}_{12}(0, 1, 1, 0)] - 4 [ \tilde{b}_{12}(0, 1, 1/\sqrt{2}, 1/\sqrt{2}) + i \tilde{b}_{12}(0, 1, 1/\sqrt{2}, i/\sqrt{2})],\label{48of64} \\
i2m_{16} &=& (1 + i) (m_{12} + m_{15} + m_{26} + m_{56}) 
- (m_{11} + m_{22} + m_{55} + m_{66}) 
+8\tilde{b}_{11}(1/\sqrt{2}, i/\sqrt{2},1/\sqrt{2}, 1/\sqrt{2})
\nonumber \\
&&\hspace{-.5cm}+8\tilde{b}_{11}(1/\sqrt{2}, 1/\sqrt{2},1/\sqrt{2}, i/\sqrt{2})\!
+\!i8[\tilde{b}_{11}(1/\sqrt{2}, 1/\sqrt{2},1/\sqrt{2}, 1/\sqrt{2}) 
\!-\!\tilde{b}_{11}(1/\sqrt{2},i/\sqrt{2},1/\sqrt{2}, i/\sqrt{2})],
\label{49of64}\\
2m_{25} &=& (1 + i) (m_{15} + m_{26}) + (1 - i)(m_{12}^* + m_{56}^*) 
- (m_{11} + m_{22} + m_{55} + m_{66}) 
+8\tilde{b}_{11}(1/\sqrt{2}, 1/\sqrt{2},1/\sqrt{2}, 1/\sqrt{2})
\nonumber \\
&&\hspace{-.5cm}+8\tilde{b}_{11}(1/\sqrt{2}, i/\sqrt{2},1/\sqrt{2}, i/\sqrt{2})\!
+\!i8[\tilde{b}_{11}(1/\sqrt{2}, 1/\sqrt{2},1/\sqrt{2}, i/\sqrt{2}) 
\!-\!\tilde{b}_{11}(1/\sqrt{2},i/\sqrt{2},1/\sqrt{2}, 1/\sqrt{2})],
\\
i2m_{38} &=& (1 + i) (m_{34} + m_{37} + m_{48} + m_{78}) 
- (m_{33} + m_{44} + m_{77} + m_{88}) 
+8\tilde{b}_{22}(1/\sqrt{2}, i/\sqrt{2},1/\sqrt{2}, 1/\sqrt{2})
\nonumber \\
&&\hspace{-.5cm}+8\tilde{b}_{22}(1/\sqrt{2}, 1/\sqrt{2},1/\sqrt{2}, i/\sqrt{2})\!
+\!i8[\tilde{b}_{22}(1/\sqrt{2}, 1/\sqrt{2},1/\sqrt{2}, 1/\sqrt{2}) 
\!-\!\tilde{b}_{22}(1/\sqrt{2},i/\sqrt{2},1/\sqrt{2}, i/\sqrt{2})],\\
2m_{47} &=& (1 + i) (m_{37} + m_{48}) + (1 - i)(m_{34}^* + m_{78}^*) 
- (m_{33} + m_{44} + m_{77} + m_{88}) 
+8\tilde{b}_{22}(1/\sqrt{2}, 1/\sqrt{2},1/\sqrt{2}, 1/\sqrt{2})
\nonumber \\
&&\hspace{-.5cm}+8\tilde{b}_{22}(1/\sqrt{2}, i/\sqrt{2},1/\sqrt{2}, i/\sqrt{2})\!
+\!i8[\tilde{b}_{22}(1/\sqrt{2}, 1/\sqrt{2},1/\sqrt{2}, i/\sqrt{2}) 
\!-\!\tilde{b}_{22}(1/\sqrt{2},i/\sqrt{2},1/\sqrt{2}, 1/\sqrt{2})],
\\
i2m_{18} &=& (1 + i) (m_{14} + m_{17} + m_{28} + m_{58}) 
- (m_{13} + m_{24} + m_{57} + m_{68}) 
+8\tilde{b}_{12}(1/\sqrt{2}, i/\sqrt{2},1/\sqrt{2}, 1/\sqrt{2})
\nonumber \\
&&\hspace{-.5cm}+8\tilde{b}_{12}(1/\sqrt{2}, 1/\sqrt{2},1/\sqrt{2}, i/\sqrt{2})\!
+\!i8[\tilde{b}_{12}(1/\sqrt{2}, 1/\sqrt{2},1/\sqrt{2}, 1/\sqrt{2}) 
\!-\!\tilde{b}_{12}(1/\sqrt{2},i/\sqrt{2},1/\sqrt{2}, i/\sqrt{2})],\\
2m_{27} &=& (1 + i) (m_{17} + m_{28}) + (1 - i) (m_{23} + m_{67}) 
- (m_{13} + m_{24} + m_{57} + m_{68}) 
+8\tilde{b}_{12}(1/\sqrt{2}, i/\sqrt{2},1/\sqrt{2}, i/\sqrt{2})
\nonumber \\
&&\hspace{-.5cm}+8\tilde{b}_{12}(1/\sqrt{2}, 1/\sqrt{2},1/\sqrt{2}, 1/\sqrt{2})\!
+\!i8[\tilde{b}_{12}(1/\sqrt{2}, 1/\sqrt{2},1/\sqrt{2}, i/\sqrt{2}) 
\!-\!\tilde{b}_{12}(1/\sqrt{2},i/\sqrt{2},1/\sqrt{2}, 1/\sqrt{2})],\\
i2m_{36} &=& (1 + i) (m_{35} + m_{46} + m_{23}^* + m_{67}^*) 
- (m_{13}^* + m_{24}^* + m_{57}^* + m_{68}^*) 
+8\tilde{b}_{12}^*(1/\sqrt{2}, i/\sqrt{2},1/\sqrt{2}, 1/\sqrt{2})
\nonumber \\
&&\hspace{-.5cm}+8\tilde{b}_{12}^*(1/\sqrt{2}, 1/\sqrt{2},1/\sqrt{2}, i/\sqrt{2})\!
+\!i8[\tilde{b}_{12}^*(1/\sqrt{2}, 1/\sqrt{2},1/\sqrt{2}, 1/\sqrt{2}) 
\!-\!\tilde{b}_{12}^*(1/\sqrt{2},i/\sqrt{2},1/\sqrt{2}, i/\sqrt{2})],\\
2m_{45} &=& (1 + i) (m_{35} + m_{46}) + (1 - i) (m_{14}^* + m_{58}^*) 
- (m_{13}^* + m_{24}^* + m_{57}^* + m_{68}^*) 
+8\tilde{b}_{12}^*(1/\sqrt{2}, i/\sqrt{2},1/\sqrt{2}, i/\sqrt{2})
\nonumber \\
&&\hspace{-.5cm}+8\tilde{b}_{12}^*(1/\sqrt{2}, 1/\sqrt{2},1/\sqrt{2}, 1/\sqrt{2})\!
+\!i8[\tilde{b}_{12}^*(1/\sqrt{2}, 1/\sqrt{2},1/\sqrt{2}, i/\sqrt{2}) 
\!-\!\tilde{b}_{12}^*(1/\sqrt{2},i/\sqrt{2},1/\sqrt{2}, 1/\sqrt{2})].
\label{64of64}
\end{eqnarray} 

Note that from Eq.~(\ref{49of64}) to Eq.~(\ref{64of64}), we have 
expressed all $m_{ij}$ partially as functions of the previous coefficients as well [Eqs.~(\ref{1of64})-(\ref{48of64})]. 
This was done to save space and because 
the full expressions for those coefficients are too cumbersome 
if fully written in terms of $\tilde{b}_{ij}$ alone.
\end{widetext}

\end{document}